\documentclass[twocolumn,preprintnumbers,amsmath,amssymb]{revtex4}
\usepackage{graphicx}

\usepackage{amsmath}
\usepackage{amssymb}
\usepackage{amsthm}
\usepackage{amsfonts}
\usepackage{enumerate}

\usepackage{centernot}
\usepackage{tikz}

\begin{document}

\title{Gauge fermions with flat bands and anomalous transport via chiral modes from breaking gauge symmetry}

\author{Xi Luo$^{1}$ and Yue Yu$^{2,3,4}$}

\affiliation {1. CAS Key Laboratory of Theoretical Physics, Institute of
Theoretical Physics, Chinese Academy of Sciences, P.O. Box 2735,
Beijing 100190, China\\
2.Center for Field
Theory and Particle Physics, Department of Physics, Fudan University, Shanghai 200433,
China \\
3. State Key Laboratory of Surface Physics, Fudan University, Shanghai 200433,
China\\
4. Collaborative Innovation Center of Advanced Microstructures, Nanjing 210093, China}

\begin{abstract}
The dispersionless longitudinal photon in Maxwell theory is thought of as a redundant degree of freedom due to the gauge symmetry. We find that when there exist exactly flat bands with zero energy in a condensed matter system, the fermion field may locally transform as a gauge field and the system possesses a 
gauge symmetry. As the longitudinal photon, the redundant degrees of freedom from the flat bands must be gauged away from the physical states.  As an example, we study spinless fermions on a generalized Lieb lattice in three dimensions. The flat band of the  longitudinal fermion induces a gauge symmetry.  An external magnetic field breaks this  gauge symmetry and emerges a bunch of non-topologically chiral modes.  Combining these emergent chiral modes with the chiral anomaly mode which is of an opposite chirality, rich anomalous electric transport phenomena exhibit and are expected to be observed in 
Pd$_3$Bi$_2$S$_2$ and Ag$_3$Se$_2$Au.
\end{abstract}

\date{\today}

\maketitle

\noindent{\it Introduction. } Recently,  the  enthusiasm of researches on the flat band (FB) is stimulated due to its importance in the topological states of matter \cite{fci4,fci5,fci6}.  The fractional quantum anomalous Hall effects were predicted  in nearly FBs \cite{fci1,fci2,fci3}. The materials supporting a topological non-trivial FB have been designed for organometals \cite{orgm}. The three dimensional FB was also experimentally observed for massless Kane fermion \cite{nphy,ncomm} and proposed in the crystals with the symmetry of space groups 199 and 214 \cite{ber}, which is pseudo-spin one analog of Weyl semimetal \cite{wan}. The interesting magnet-optics of massless Kane fermions was studied \cite{magop,magop1}.  

 The tight-binding models on specific lattices lead to  exactly FBs \cite{Lieb,aoki,Deng,Miy,WuCJ,Berc}. They were designed and  experimentally realized in cold atoms on optical lattices. A photonic crystal Lieb lattice has been experimentally demonstrated \cite{photonic1,photonic2}. It may also be simulated by circuit quantum  electrodynamics simulator \cite{WuYin}. 
 Because of the infinite degeneracy of the FBs and  the singular density of states at Fermi surface, many astonishing new phenomena may emerge. Wigner crystals \cite{WuCJ} and superfluid \cite{volovik, Zhai,Julku} become possible at high temperature. Exotic heavy excitons may exist \cite{Chamon}. In an external magnetic field, Landau levels of charged fermions are proportional to the square root of odd integers 
 while there are huge number of FBs corresponding to Landau indices \cite{Goldman}.  A large incident angle Klein tunneling was predicted \cite{Xing}. For Dirac-Weyl fermions with integer spin, it was asserted that there is no chiral anomaly because the Dirac-like operator is not  an elliptical operator \cite{Lan}. The  dc conductivity will diverge due to the FB \cite{Vigh}.  

In this Letter, we would like to point out that the exact zero FBs might not be the physical degrees of freedom (DOF). We will first recall the Maxwell theory in which the longitudinal (LG) photon is dispersionless but unphysical. The spin-1 fermion is of the same gauge symmetry and has the same LG component which is flat and must also be gauged away.  Many results based on the FB DOF have to be reconsidered. To recover these lost DOF, one must break the gauge symmetry. These restored gauge DOF  will cause interesting consequences. 

We will study the spinless fermions on a generalized Lieb lattice in three dimensions. By choosing the hopping constants, there are FBs corresponding to the gauge symmetry. The fermions share the properties both of a gauge field and a matter field. We call this gauge symmetry the Bloch FB gauge(.
Focusing on the fermion with a linear dispersion, one finds that in an external magnetic field, the BFBG symmetry is broken. The Landau level spectra are exactly calculated. Besides a chiral mode in the lowest Landau level which arises from the chiral anomaly \cite{NN}, there are many chiral modes with opposite chirality across the zero energy level. Each one corresponds to a given Landau index $n>1$ and these chiral modes stem from the breaking of the BFBG symmetry. Many chiral modes contribute to the conductance in a small applied electric field $E$  parallel to the magnetic field so that the magnetoresistance(MR) approaches $-1$.  As  $E$ increases, the number of the chiral modes participating the transport decreases. This leads to a giant positive MR first and then a crossover to a giant negative MR.  Finally,
all chiral modes but the topological one quit the transport, which gives a large negative MR similar to that in the Weyl semimetals.   Since this kind of pseudo-spin one models have its material correspondence, e.g., 
Pd$_3$Bi$_2$S$_2$ with the symmetry group 199 and Ag$_3$Se$_2$Au with the symmetry group 214 \cite{ber}, we anticipate  these anomalous transport
phenomena can be experimentally observed. 
\\

\noindent{\it "Primordial FBs" in gauge fields.} The history of researching on the FBs  in condensed matter systems  casts back to 1970s in amorphous semiconductors \cite{wea,weath,thwea}.  The FB means there are extremely strong correlations in  many body  systems \cite{esc1,esc2}.
There was a much earlier physical theory, the Maxwell theory of the electromagnetic field, in which the LG photon is dispersionless and so is a "primordial FB"!  The Maxwell equations read
$
(\partial^2g_{\mu\nu}-\partial_\mu\partial_\nu)A^\nu=0
$.
Taking $A_0=0$,  the equations of motion of $A_i$  are given by 
 \begin{equation}
		( -\nabla^2\delta_{ij}+\partial_i\partial_j)
	A_j=E^2A_i. \label{MW}
\end{equation}
We here have not yet considered the residual gauge symmetry. 
It is easy to check that $A^{(0)}_\mu({\bf r})=(0,\partial_i \Lambda({\bf r}))$ with  arbitrary  time-independent scalar function $\Lambda({\bf r})$ is a zero energy solution. In the momentum space, $A^{(0)}_\mu({\bf q})=(0,q_i\Lambda({\bf q}))$ is an exactly FB. 
This zero energy FB is identified as the dispersionless LG photon which is unphysical redundant DOF.  In fact, $(0,\partial_i \Lambda({\bf r}))$ is the residual time-independent gauge transformation in $A_0=0$ gauge.

A massless higher-spin matter field can also transform as a gauge field \cite{RS}.  For example, for spin-$\frac{3}2$ relativistic  spinor-vector $\Psi_{\nu\alpha}$ \cite{note},  Rarita and Schwinger in 1941 recognized that the Lagrangian 
$
L_{\rm RS}=\bar{\Psi}_\mu((g^{\mu\nu}\gamma \cdot \partial)-(\gamma^\mu\partial^\nu+\partial^\mu\gamma^\nu)+\gamma^\mu(\gamma\cdot\partial)\gamma^\nu)\Psi_\nu
$
is  invariant under the gauge transformation $\Psi'_{\mu\alpha}=\Psi_{\mu\alpha}+\partial_\mu\eta_\alpha$ \cite{RS}. Therefore, the gauge invariance under a static gauge transformation $\delta \Psi^{(0)}_{\mu\alpha}=(0,\partial_i \eta_\alpha({\bf r}))$  implies  that $\delta \Psi^{(0)}_{\mu\alpha}({\bf q})$ correspond to  four zero energy FBs.  These FBs are redundant DOF.  The genuine physical DOF are the Rarita-Schwinger-Weyl fermions with the highest and lowest helicity \cite{Lure}. In fact, any higher spin relativistic massless particles are of such a gauge symmetry \cite{RS}. The physical DOF are two Rarita-Schwinger-Weyl fermions \cite{Lure}.  
\\

\noindent{\it FB in spin-1 gauge fermions. } Keeping the redundant FBs of the LG  photon and gauge fermions in mind, we exam other possible gauged FBs. The pseudo-spin one fermions are frequently investigated \cite{Lieb,Berc}. In three dimensions, the Hamiltonian is given by \cite{ber}
 \begin{eqnarray}
 H_{\rm s1}={\bf q}\cdot {\bf S}, \label{s1}
 \end{eqnarray}    
where ${\bf S}$ are $3\times3$ matrices of SO(3) generators. For convenience,  we take $S^i_{jk}=-i\epsilon_{ijk}$. The solutions of the Schr\"odinger equations are well-known: There are two bands $\boldsymbol{\psi}^{(\pm)}$ with the eigen energies $\pm q$ and a FB $\boldsymbol{\psi}^{(0)}=\frac{\bf q}q$. However, it is easy to check that the Schr\"odinger equation $H_{ij}\psi_j=E\psi_i$ is gauge invariant under $\delta \boldsymbol{\psi}={\bf q}\Lambda({\bf q})$ which is the FB solution. In the spirit of gauge symmetry, the FB solution is redundant and must be gauged away! Actually,  if we define the static transverse (TS) projection $P_T=\delta_{ij}-\frac{q_iq_j}{q^2}$, the FB is the LG fermion while $\boldsymbol{\psi}^{(\pm)}$ are the TS fermions, similar to the electromagnetic field. The gauge DOF $\delta \boldsymbol{\psi}={\bf q}\Lambda({\bf q})$ belong to the space of states $P_L\psi=\psi_L$ with $P_L=\frac{q_iq_j}{q^2}$. Namely, this spin-1 fermion is a gauge field. To distinguish with the electromagnetic gauge symmetry, we call this gauge symmetry the BFBG symmetry. This BFBG fermion may be realized in 
the materials with the symmetry of space groups 199 and 214 \cite{ber}.\\

\begin{figure}[ptb]
\centering
\includegraphics[width=7.5cm]{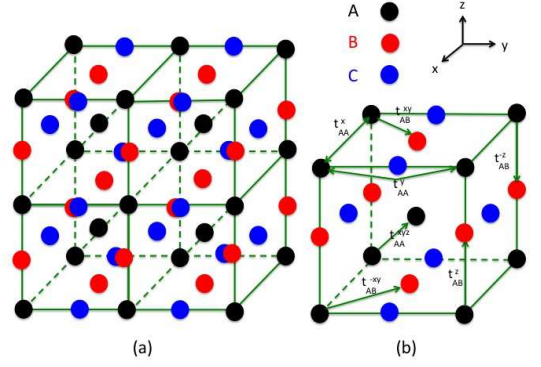}\newline\caption{ (Color online) The three-dimensional generalized Lieb lattice. (a) Lattice structure with three body-centered cubic lattices coupled together. (b) The hopping constants set. The others are similar.  We take  $t^z_{AB}=t_{AC}^y=t_{BC}^x=-t^{-z}_{AB}=-t_{AC}^{-y}=-t_{BC}^{-x}=t_1$; $t^{xy}_{AB}=t^{-x,-y}_{AB}
=t_{AC}^{xz}=t_{AC}^{-x,-z}=t_{BC}^{yz}=t_{BC}^{-y,-z}=-t^{-xy}_{AB}=-t^{x,-y}_{AB}
=-t_{AC}^{-xz}=-t_{AC}^{x,-z}=-t_{BC}^{-yz}=-t_{BC}^{y,-z}=t_2$; $t_{AA}^{xyz}=t_{BB}^{xyz}= t_{CC}^{xyz}=t_3$; $t_{AA}^y=t^z_{AA}=t_{BB}^x=t^z_{BB}=t_{CC}^x=t^y_{CC}=t_4$; $t_{AA}^x=t_{BB}^y=t^z_{CC}=t'_4$,  and so on.}%
\label{fig1}%
\end{figure}

\noindent{\it Generalized Lieb lattice. } Furthermore, we can show that a general FB across the Fermi level is corresponding to a local BFBG transformation and thus is redundant. For example, we study a generalized Lieb lattice model. 
The tight-binding model of spinless fermions includes the first, second, third and fourth neatest neighbor hoppings. (See Fig. \ref{fig1}.) We take $t_3=0$ for simplicity.  After adjusting the chemical potential and shifting the zero energy to the Fermi level, the Hamiltonian is then given by
\begin{eqnarray}
H_{\rm GL} &=&2t_1\boldsymbol{\kappa}\cdot{\bf S}
+2t_4\kappa^2 +2(t'_4-t_4)(\kappa^2_i\delta_{ij})_{3\times 3}\nonumber\\
&-&2t_2 (\delta_{ij}\kappa_i^2-\kappa_i\kappa_j)_{3\times3}, \label{GL}
\end{eqnarray}
where $\boldsymbol{\kappa}=(\sin q_x, \sin q_y,\sin q_z)$. Obviously, at the $\Gamma$ point, the first term in $H_{\rm GL}$ recovers (\ref{s1}). 

We consider two special cases with FBs. First, we take $t_2=-t_4,t_4'=0$, the Hamiltonian (\ref{GL}) is reduced to
\begin{eqnarray}
H^T_{\rm GL}=2t_1\boldsymbol{\kappa}\cdot{\bf S}+2t_4(\kappa^2\delta_{ij}-\kappa_i\kappa_j)_{3\times3},
\end{eqnarray}
where the second term at the $\Gamma$ point is just the Maxwell Hamiltonian in (\ref{MW}) but acting on the fermons. Then, the TS Bloch modes $\boldsymbol\psi_\pm=\frac{1}{\sqrt2 \kappa}((\kappa_x\kappa_z\mp i\kappa\kappa_y)/\sqrt{\kappa_x^2+\kappa_y^2},(\kappa_y\kappa_z\pm i\kappa\kappa_x)/\sqrt{\kappa_x^2+\kappa_y^2},-\sqrt{\kappa_x^2+\kappa_y^2})$ have the dispersions $E_\pm=\pm 2t_1 \kappa-2t_4\kappa^2$. The LG mode $\boldsymbol\psi_0= \boldsymbol\kappa\Lambda({\bf q})$ is a flat Bloch band lying on the Fermi level, which is redundant and must be gauge away. This is the TS fermion model. 
  
  The second case is to take $t_1=t_4=0$ and $t'_4=-t_2$ and then the Hamiltonian (\ref{GL}) is reduced to the LG part 
 \begin{eqnarray}
 H^L_{GL}=-2t_2(\kappa_i\kappa_j)_{3\times3}.\label{LH}
 \end{eqnarray}
 The LG mode has a dispersion $E_0=-2t_2\kappa^2$ while two TS modes $\boldsymbol{\psi}_\pm\Lambda_\pm({\bf q})$ are FBs which must be gauged away.  This is the LG fermion BFBG model.

 
 The model (\ref{GL}) can also be reduced to two dimensions. 
 There were plenty of works in this Lieb lattice model as mentioned in {\it Introduction}.  The Hamiltonians in those works are similar to the two-dimensional reduction of (\ref{s1})  up to a unitary transformation.  A similar model can also be realized on ${\cal T}_3$ \cite{Berc}. 
 
\noindent{\it Comments on the physical observables and nearly FB. } 
All physical properties related to the FB DOF have to be reexamined. The infinite degeneracy and then the singular density of states are unphysical.   The particle number and currents, although they are conserved quantities, are not BFBG invariant and so cannot be physical observables. A BFBG invariant observable can be obtained by replacing the fermion field $\psi$ in the operator with its physical space projection, say,  $P_T\psi$ or $P_L\psi$ for the TS  or LG model, respectively.  It was asserted that the chiral anomaly does not exist in the spin-1 system with the Hamiltonian (\ref{s1}) because the corresponding "Dirac"  operator is the curl operator which is not elliptic. However, if the FB is removed, the zero modes in the physical space are finite. In fact, the TS fermions carry helicity $\pm1$ which give the Chern numbers $\pm2$ if $q_3\ne 0$ is fixed. The chiral anomaly is given by the Chern numbers and there is a quantum anomalous Hall effect with $C=2$ in two-dimensions. The Fermi arcs on the surface of the system exist \cite{ber}, similar to that on the surface of the Weyl semimetal \cite{wan}. The chiral anomaly effect will also be seen in the next section when we discuss the model in the external magnetic field.    

Because the FB concerned here exactly lies in the Fermi level, the finite energy behaviors studied before may still be correct, e.g., the peculiar Klein tunneling for the spin-1 fermion \cite{Xing}.

We do not yet study the nearly FBs. There are several situations:(i) If the nearly FB can be decomposed into an exactly flat part plus a perturbation, the exactly flat DOF are still gauged ones, i.e., the unperturbed Hamiltonian is the BFBG invariant one and the BFBG is only perturbatively broken.  
(ii) There is no such a decomposition for the nearly FB. For example, if the nearly FBs  in two dimensions are topologically nontrivial, they cannot  continuously tend to exactly flat \cite{CLi}. However, in three dimensions, there are topologically nontrivial exactly FBs. The TS modes for the Hamiltonian (\ref{LH}) are examples. They are exactly flat but their monopole charges are $\pm2$. Reducing to two dimensions, $\boldsymbol\psi_\pm$ are topologically trivial, in consistent with the theorem proved in \cite{CLi}. (iii) The BFBG symmetry is nonperturbatively broken. For instant, an external electromagnetic field is applied, as we will study right below.    
 
 \noindent{\it Landau levels. }  The large negative MR  is a ubiquitous property in topological insulators, Dirac and Weyl semimetals, stemming from the chiral anomaly \cite{NN}.  These phenomena were observed experimentally \cite{Kim,LiSY,Huang}. In this BFBG fermion system, there may be such an anomalous electric transport phenomenon in the lowest Landau level due to the chiral anomaly \cite{ber}. Moreover, we will show that the electric transport phenomena  for the psuedo-spin 1 system in an external magnetic field system are much richer than in the spin-1/2 systems.

\begin{figure}[ptb]
  	\centering
  	\includegraphics[width=7.8cm]{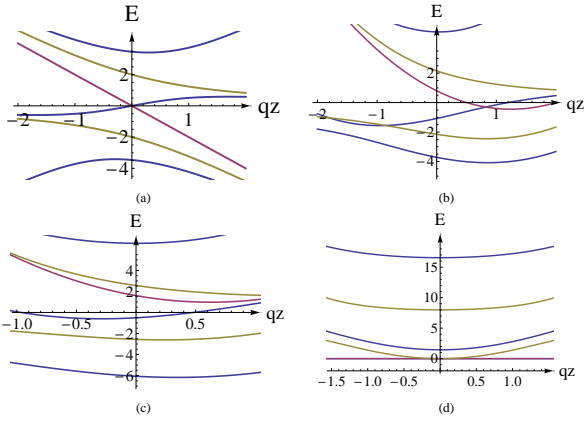}\newline\caption{ (Color online) The purple, brown and blue curves represent the lowest, the second  and the third Landau levels. The parameters are chosen as: (a) $t_1=1, t_2=t_4=0$, (b) $t_1=1, -t_2=t_4=0.4$, (c) $t_1=1, -t_2=t_4=0.8$, (d) $t_1=t_4=0, t_2=-1$.}
  	\label{fig2}%
  \end{figure}

 We consider the long wave length of the Hamiltonian (\ref{GL}) for the charged fermions in an external  magnetic field  applied in the $z$ axis. For simplify, we take  $t_4'-t_4=t_2$.  Due to the breaking of the translational symmetry in the $x$-$y$ plane, we work in the real space. The Hamiltonian at $\Gamma$ point then reads 
 \begin{eqnarray}
 H_{M}=2t_1{\bf S}\cdot{\bf D}+2t_4D^2+2t_2(D_iD_j)_{3\times3}, \label{EM}
 \end{eqnarray}
where $D_a=-i\hbar\partial_a+\frac{e}c A_a$ ($a=1,2$) with the magnetic field ${\bf B}=\nabla\times {\bf A}$ in the $z$ direction while $D_3=q_z$ since $A_3=0$. We take $\hbar=e=c=1$ and the magnetic length $l_B=\sqrt{1/B}=1$. Due to $[D_1,D_2]=iB=i$, the Hamiltonian (\ref{EM}) is not BFBG invariant under $\delta\boldsymbol{\psi}={\bf D}\Lambda$ even if $t_2=-t_4$.  (Do not confuse with the electromagnetic gauge transformation.) This Hamiltonian can be analytically diagonalized.  To see this, we make a basis rotation $\Phi_{1,3}=\frac1{\sqrt2}(i\psi_2\pm \psi_1), \Phi_2=\psi_3$,  and transform the Hamiltonian $H_M$ to
\begin{eqnarray}
 \left(
\begin{array}{ccc}
	I_1
	& (2t_1+2t_2q_z)a^\dagger
	&-2 t_2a^{\dagger 2} \\ 
	(2t_1+2t_2q_z)a
	& I_2
	& (2t_1-2t_2q_z)a^\dagger \\ 
	-2 t_2a^2
	& (2t_1-2t_2q_z)a
	& I_3
\end{array}
\right),\label{TH}
\end{eqnarray}
where $a$ and $a^\dag$ obeying $[a,a^\dag]=1$ are the Landau level lowing and raising operators;
$I_1=(4t_4+2t_2)a^\dagger a+2t_4-2t_1q_z+2t_4q_z^2$, $I_2=4t_4a^\dagger a+2t_4+(2t_4+2t_2)q_z^2$ and $I_3=(4t_4+2t_2)a^\dagger a+(2t_4+2t_2)-2t_1q_z+2t_4q_z^2$. Denoting the $n$-th Landau level wave functions by $\phi_n$, general solutions of the problem are $\boldsymbol{\Phi}_n=(C_1\phi_n,C_2\phi_{n-1},C_3\phi_{n-2})$ where $\phi_{-1,-2}$ are taken to be 0. For $n=0$, there is only one Landau band with $E_0=2t_4-2t_1q_z+2t_4q_z^2$, which is a chiral mode in a chain along the $z$-direction when $t_1^2-4t_4^2>0$ (See Figs. \ref{fig2}(a) and (b)) and comes from the chiral anomaly as that for the Weyl semimetal. The chiral anomaly is destroyed by the larger quadratic term if $t_1^2-4t_4^2<0$. See Fig. \ref{fig2}(c) where the chiral mode does not cross zero energy.  For $n=1$, there are two Landau bands as shown in Fig. \ref{fig2} and they also relate to the chiral anomaly \cite{beenkker}.  

For a given $n>1$, three Landau bands are obtained by solving the eigen equation of (\ref{TH}). For $t_2=t_4=0$ and a given $n$, two bands are gapped with energies which are $\pm2t_1\sqrt{2n+1}$ at $q_z=0$ and there is a gapless chiral mode whose chirality is opposite to the chiral mode at the lowest Landau level. The slope of the gapless chiral  band at $q_z=0$ is $\frac{2t_1}{2n+1}$. The chiral mode is not topologically protected but stems from the breaking of the BFBG symmetry. Fig. \ref{fig2}(a) shows the spectra of the  first, second and third Landau levels for this case. Fig. \ref{fig2}(b) depicts the results with the quadratic perturbation, which are similar to $t_2=t_4=0$. As the quadratic terms dominate, the chiral modes become non-chiral as shown in Fig. \ref{fig2}(c).  

For the LG gauge fermion model ($t_1=0$, $t_2=t_4'$ and $t_4=0$), there is only an $E=0$ band for $n=0$. For $n=1$, two gapped bands are given by $E^2-2t_2(2+q_z^2)E+2t_2q_z^2=0$. For $n>1$, there is always an $E=0$ band and the other gapped two are given by $E_{n,\pm}=(2n+q_z^2)t_2\pm t_2\sqrt{4(n^2+n-1)+4nq_z+5q_z^2}$. These dispersions are plotted in Fig. \ref{fig2}(d).  Notice that the FBs here cannot be gauged away because timing an arbitrary function, the FB wave functions are no longer flat.       
Due to the space limit, we do not investigate this LG gauge fermion further. 
   
   \begin{figure}[ptb]
\centering
\includegraphics[width=8cm]{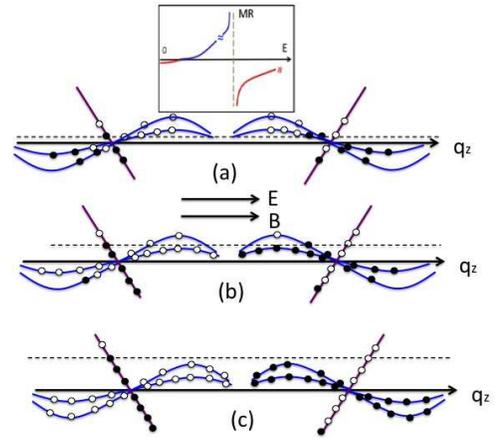}\newline\caption{(Color online) The quantum charge pumping between Weyl nodes in parallel electric and magnetic fields. Only the gapless modes are shown. The dashed lines are the chemical potential of the left-moving fermions due to the variation of the $E$ field. The purple lines are the lowest Landau level chiral modes while the blue curves are the chiral modes of higher Landau levels. (a) In a very weak $E$ field, a large amount of chiral modes contribute to the electric current. (b) A few modes, say between 10 to 20, contribute. (c) Only the lowest Landau level chiral modes do. The inset is the schematic of the MR.}
\label{fig3}%
\end{figure}

\noindent{\it The magnetic transport property.}  The giant negative MR is the significant phenomenon in topological semimetals \cite{qi,yan}.   Due to the existence of the new emergent chiral modes, combining with the topological chiral mode, the transport phenomena are much richer. We focus on the model with the spectrum of Fig. \ref{fig2}(a) and consider the transport around the zero energy. For a given $n>1$, the energy of the chiral mode is of the maximum and minimum $\sim\pm t_1/\sqrt {n+1/2}$ and tends to zero as $1/q_z$ when $|q_z|\to \infty$. We notice that all these chiral modes are orthogonal. Especially, the backward scattering between the anomaly mode and the other modes from the BFBG breaking is forbidden.   We consider the quantum charge pumping between two Weyl points \cite{qi}, which are, e.g., the $\Gamma$ and $(0,0,\pi)$ points here. First, apply a very weak electric field $E$ along the $z$-direction so that the chemical potential difference $\Delta\mu$ between left- and right-moving fermions is of the order $t_1/\sqrt {n_{max}}$ with a very large $n_{max}$.  The $n_{max}$ chiral modes as well as the anomaly mode participate the charge transport. (See Fig. \ref{fig3}(a)). The reason that the chiral modes with $n>n_{max}$ do not contribute to the current is as follows: As seen in Fig. \ref{fig3}(b), when a chiral mode with its maximal energy is smaller than $\Delta\mu/2$, all charges in the left Weyl point are pumped to the right Weyl point. Since the filled chiral band includes both states with positive and negative velocities, the contributions of all states to the current cancel.  The current is given by 
\begin{eqnarray}
j_z\propto -2t_1(1-\sum_{n=2}^{n_{max}}\frac{1}{2n+1})\sim t_1\ln n_{max},
\end{eqnarray}
where the right side of $\sim$ is the result for a large $n_{max}$.
This means the resistivity $\rho(B)$ is much smaller than $\rho_0$, the zero field resistivity. The MR $(\rho(B)-\rho_0)/\rho_0\sim -1$. As $E$ becomes stronger so that the chemical potential difference is raised to, e.g., $t_1/\sqrt {n_{max}+1/2}$ with  $n_{max}=14$,
$j_z\propto 0.005t_1$, which is much smaller than the magnitude of the anomaly mode contribution which is $\propto -2t_1$ and then gives a positive MR whose magnitude is 400 times larger than the negative MR from the chiral anomaly.  Reducing $n_{max}$ to $13$, $j_z\propto -0.063t_1$.  The negative MR is 30 times large than that from the chiral anomaly. Finally,  when $\Delta\mu/2$ is larger than the maximum of the $n=2$ chiral mode energy (see Fig. \ref{fig3}(c)), all chiral modes but the anomaly one do not contribute. The negative MR is the same as that in Weyl semimetal \cite{beenkker}. The scketch of the MR of $E$ is shown in the inset of Fig. \ref{fig3}.  \\

 \noindent{\it Conclusions.}  We have studied the gauge fermions with FBs.  We pointed out that the exactly FBs in a band theory are the gauge redundant DOF, similar to the LG photon in Maxwell theory.  We argued that all results involving in the redundant DOF must be reconsidered. The external magnetic field breaks the BFBG and the redundant DOF are released. Due to the existence of them, the transport phenomena in the magnetic field become very plentiful. Since the model we studied has the material correspondence, say, 
 Pd$_3$Bi$_2$S$_2$ and Ag$_3$Se$_2$Au \cite{ber}, we expect these phenomena are experimentally observable.

We thank Gang Chen and Fei Teng for helpful discussions. This work was supported by NNSF of China (11474061).

\end{document}